\documentclass[aps,prl,twocolumn,superscriptaddress,showpacs,
               amsmath]{revtex4}

\usepackage{dcolumn}
\usepackage{bm}
\usepackage{amsfonts}
\usepackage{graphicx,graphics}
\usepackage{amssymb}
\usepackage{rotate}
\usepackage{subfigure}
\usepackage{epsf}
\usepackage{epsfig}
\usepackage{amsmath}
\usepackage{dcolumn}

\begin{document}

\title{\textbf{A Three Dimensional Kasteleyn Transition: Spin Ice in a $[100]$ Field}}

\author{L. D. C. Jaubert}
\affiliation{Laboratoire de Physique, \'Ecole Normale Sup\'erieure de Lyon, Universit\'e de Lyon, CNRS, 46 All\'ee d'Italie, 69364 Lyon Cedex 07, France.}

\author{J. T. Chalker}
\affiliation{Theoretical Physics, Oxford University, 1
Keble Road, Oxford, OX1 3NP, United Kingdom.}

\author{P. C. W. Holdsworth}
\affiliation{Laboratoire de Physique, \'Ecole Normale Sup\'erieure de Lyon, Universit\'e de Lyon, CNRS, 46 All\'ee d'Italie, 69364 Lyon Cedex 07, France.}

\author{R. Moessner}
\affiliation{Theoretical Physics, Oxford University, 1
Keble Road, Oxford, OX1 3NP, United Kingdom.}
\affiliation{Max-Plack-Institut f\"ur Physik komplexer Systeme, 01189 Dresden, Germany}

\date{\today}

\begin{abstract}
We examine the statistical mechanics of spin-ice materials with a $[100]$ magnetic field. We show that the approach to saturated magnetisation is, in the low-temperature limit, an example of a $3d$ Kasteleyn transition, which is topological in the sense that magnetisation is changed only by excitations that span the entire system. We study the transition analytically and using a Monte Carlo cluster algorithm, and compare our results with recent data from experiments on Dy$_2$Ti$_2$O$_7$.
\end{abstract}


\maketitle

Topological phases occur in connection with new liquid states and
concurrent exotic transitions out of them. A central idea is the
emergence of degrees of freedom which cannot be defined or manipulated
locally. Their presence has profound ramifications for the static and dynamic properties of such
phases. They can exhibit algebraic correlations in the absence of
criticality~\cite{Youngblood} and unusually slow relaxation towards
thermal equilibrium~\cite{Henley,Kondev}.

\begin{figure}
\includegraphics[scale=0.51]{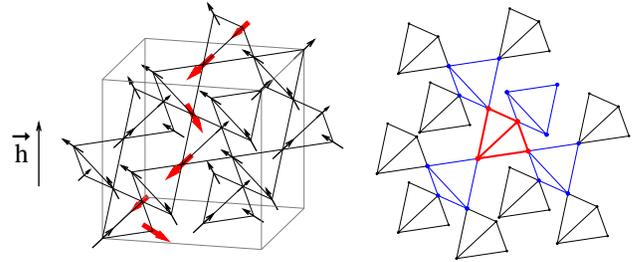}
\caption{(Color online) \textit{Left}: Pyrochlore lattice showing $q=0$ spin structure (black
narrow spins) and a string defect (red thick spins).
\textit{Right}: Bethe lattice with a central tetrahedron (red) and
first layer of nearest-neighbour tetrahedra (blue).} \label{pyrochlore-spin}
\end{figure}

A magnetic system whose topological properties have a particularly
simple and transparent origin is known as spin
ice~\cite{Anderson56,Harris97} (Fig.~\ref{pyrochlore-spin}): in its
macroscopically degenerate ground state, a magnetic version of the ice rules apply. These
stipulate that for each tetrahedron of the corner-sharing pyrochlore
lattice, two of the four spins point in, and two point out of the
tetrahedron along the local body centred diagonal axes.  As a result,
the lattice divergence of the spin field vanishes everywhere:
$\nabla\cdot\vec{S}=0$. As a consequence of this local constraint, the
magnetisation of each $(100)$ plane is the same~\cite{Moessner98} and is
therefore a topological quantity which can be changed only by making a
simultaneous change throughout the system: the smallest such
magnetic excitation involves a set of spins on a string spanning
the system. 
An
attractive feature of spin-ice materials is that one can couple
directly to this topological quantity using a
uniform magnetic field, which lifts the macroscopic degeneracy of the
zero-field ground states.

In this paper, we focus on the behaviour near saturation, when the
ordered $q=0$ state~\cite{Harris97,Ramirez99} is reached by applying a magnetic field
along the $[100]$ direction as illustrated in
Fig.~\ref{pyrochlore-spin}. In the saturated state at large fields,
the ice rules prohibit local fluctuations in the limit of low
temperature $T$. The resulting phase transition as the field is lowered is a three dimensional
($3d$) example of a Kasteleyn transition~\cite{Bhattacharjee83}, first proposed in
the context of $2d$ dimer models~\cite{Kasteleyn63}. Its hallmark is an
asymmetric character, appearing to be first-order on one side, and
continuous on the other. The Kasteleyn transition has in the past been
associated with soft matter: the trans-gauche disordering transition
for polymers embedded in lipid bilayer system, observed via density
measurements in dipalmitoyl lecithin (DPL)~\cite{Nagle73}. 
To our knowledge this is the first example of a three-dimensional 
Kasteleyn transition in a magnetic system.

We analyse the transition using a combination of analytical and
numerical tools. By mapping the strings onto bosonic
worldlines, we obtain an analytical expression for
thermodynamic quantities close to the critical point, and for
correlation functions at all values of the magnetic field. These predictions
are confirmed to high precision by numerical
simulation using a Monte Carlo algorithm incorporating a non-local
cluster update. In addition, we find that a calculation on a Bethe
lattice provides an excellent description of the thermodynamics of
the magnet away from the critical point. A crucial characteristic of
this transition is that it is symmetry-sustaining, as recognised in
Ref.~\cite{Harris98}; the first-order liquid-gas
transition reported there, however, is an artefact of the local
dynamics employed at that time. We compare our results with
magnetisation measurements on the spin ice compound
Dy$_2$Ti$_2$O$_7$~\cite{Fukazawa02}: they
represent well the observed behaviour at large
field and temperature, but for lower values the experimental system falls
out of equilibrium.

We consider a spin ice model with nearest-neighbour
magnetic interactions only. The Hamiltonian reads
\begin{equation}\label{Hamiltonian}
H= -J \sum_{\langle ij \rangle} \vec S_i . \vec S_j - \vec h .
\sum_i \vec S_i\,,
\end{equation}
where $\vec S_i$ is a vector of unit length, encoding an Ising spin
constrained to lie along the local body centred cubic axes of the
tetrahedron; $J>0$
is the ferromagnetic exchange constant and $\vec h$ is an external
field of strength $h$ in reduced units, aligned along the $[100]$
direction. This model is a simplification in that the real
Hamiltonian is known to incorporate dipolar interactions of at
least equal importance~\cite{Siddharthan99,Hertog00}. However, it
has been demonstrated that a nearest-neighbour description is
adequate {\em as long as the ice rules are not violated}, since
corrections
fall off with separation $r$ as $r^{-5}$ \cite{Isakov05}. Our
results should therefore be robust for $T$ $\ll J$, when behaviour
depends only on the ratio $h/T$, whereas interactions 
between further neighbours may become important
at higher $T$. 

Useful insights into the Kasteleyn transition come from discussing
in detail the string excitations introduced above: start from the
$q=0$ state, flip spins in successive $[100]$ layers to generate
another configuration, and link flipped spins in adjacent layers
with segments of the strings. This mapping
between spin and string configurations is one-to-one
provided we adopt a fixed convention for the paths followed by two strings through the same tetrahedron. Strings of finite length do not occur in ice-rule states as their
ends cost exchange energy ${\cal O}(J)$. In contrast, strings that
span the system cost only Zeeman energy, of $2h/\sqrt{3}$ per
segment. 
An isolated string also has an
entropy, of $\ln2$ per segment, and a finite density of strings can
account for the macroscopic ground state entropy of spin ice.
The total free energy
for a single spanning string of $L$ segments is
\begin{equation}\label{Kasteleyn}
{G}= L \left({2h/{\sqrt{3}}}-T\ln2\right).
\end{equation}
From this one obtains the {\it exact} value of the Kasteleyn
transition temperature, $T_{\rm K} = 2h/\sqrt{3} \ln2$, and for $h,T
\ll J$, properties are a function only of $T/T_{\rm K}$.  For
$T<T_{\rm K}$ the equilibrium phase is the $q=0$ state without any
thermal fluctuations, while for $T>T_{\rm K}$ strings are
present. Depending on their interactions, different scenarios
are possible for the variation of their density, and hence sample
magnetisation, with $T/T_{\rm K}$. If strings attract, their density
has a jump at the transition. At finite string density, if they form a commensurate crystal,
there is a magnetisation plateau; and if they form a liquid, their
magnetisation varies smoothly. In fact, by construction, strings have hard
core repulsion, and the simulations we report below show only a liquid
phase.
%

By thinking of the strings as world lines for hard core bosons moving in two
dimensions at zero temperature, with the $[100]$ direction as
imaginary time, we can obtain thermodynamic behaviour near the
critical point. Under this mapping, the reduced temperature difference $t
\equiv (T/T_{\rm K}-1)$
is proportional to the chemical potential for bosons, the
deviation of the magnetisation $M$ per site from its saturation value $M_{\rm
  sat}$ is proportional to boson density, and the spin-ice free energy
is proportional to the boson ground-state energy. The Kasteleyn
transition is between a vacuum state and a Bose condensate as the
chemical potential changes sign. Here, three (2+1) is the upper critical dimension,
since in the quantum description the dynamical critical exponent takes the value $z{=}2$, a reflection of the fact that the $[100]$ field renders the classical model highly
anisotropic. Transcription of
established results for the boson problem~\cite{Fisher88} gives
for $t>0$
\begin{equation}
M = M_{\rm sat} [ 1 - at - bt\ln(1/t)]\,,
\label{magn}
\end{equation}
where $a$ and $b$ are constants. It follows from this that the
heat capacity $C_{\rm h}$ and the differential susceptibility
$\chi$ diverge logarithmically in $t$ as the critical point is approached from
the high-temperature side.  Similar conclusions for a dimer model were reached
using arguments based on the interaction between two
strings~\cite{Bhattacharjee83,Nagle89}.

Correlation functions are of great interest throughout the
partially magnetised phase, and at $h=0$ are known to have an
algebraic dipolar form that is a consequence of the ice
rules~\cite{Isakov04}. The bosonic description of the system that
we have introduced provides both a new way of viewing this result
and also a means of finding how it changes with $h$. For a complete
treatment, it is necessary to take into account the distinction
between the four sites in the basis of the pyrochlore lattice when constructing
the mapping between spins and strings. Moreover, a precise
representation of the transfer matrix for strings in terms of a boson
Hamiltonian would include multiparticle interactions, which
are not expected to influence long-distance behaviour or critical
properties, since both of these should be universal.
We defer these details to a future publication and present here a simplified calculation, in
which bosons move on a square lattice and the imaginary time
direction is approximated as continuous. We start from a boson
Hamiltonian written in terms of spin $S=1/2$ operators as
\begin{equation}
{\cal{H}}= -
{\cal{J}}\sum_{\langle ij \rangle}({\cal{S}}^x_i{\cal{S}}^x_j+{\cal{S}}^y_i{\cal{S}}^y_j)
- {\cal B} \sum_i {\cal S}^z_i\,,
\label{XY-Hamiltonian}
\end{equation}
where ${\cal{J}}$ is an effective coupling constant and ${\cal B}
\propto h$. In this incarnation, the Kasteleyn transition is a
quantum phase transition in which the two-dimensional $\rm XY$
ferromagnet is fully polarized by a field $\cal B$ coupled to the
$z$-component of magnetisation, at a critical field strength
${\cal B}={\cal B}_{\rm  c}$. Moreover, $M$ for spin ice is
proportional to $\langle {\cal S}_i^z\rangle$ for the quantum
magnet.
We use a large $S$ approximation to evaluate the
connected correlation function, defined for spin ice as
\begin{equation}
C({\bf r},z) = \langle \vec S_i \cdot \vec S_j \rangle -
\langle \vec S_i
\rangle \cdot \langle \vec S_j \rangle \,,
\end{equation}
where the separation between lattice sites $i$ and $j$ is $\bf r$ 
within the $(100)$ plane and $z$ along
the $[100]$ direction. Since we have treated the lattice structure
in a simplified way, our results apply only to intra-sublattice
correlations. Writing ${\cal B}/{\cal B}_{\rm c} = \cos\theta$,
we find for large $r$ and $z$
\begin{equation}
C({\bf r},z) = {S\sin\theta \over{8\pi}} {3[z\sin\theta ]^2 -
(r^2+[ z\sin\theta ]^2)\over{(r^2+[ z\sin\theta ]^2)^{5/2}}}\,.
\label{correlation-function}
\end{equation}
For ${\cal B}=0$ the form of this expression reduces to that of
Ref.~\cite{Isakov04} and has the cubic symmetry of the zero field
problem. For finite field this symmetry is broken and distances in the
$z$ direction acquire a scale factor $1/{\sqrt{1-({\cal B}/{\cal B}_{\rm c})^2}}$.
Close to the Kasteleyn transition, there is a second regime of
behaviour, which also follows from Eq.~(\ref{XY-Hamiltonian}) but
is not displayed in Eq.~(\ref{correlation-function}). It arises
because correlations at distances shorter than the string
separation are controlled by the behaviour of an isolated string,
and hence reduce to the correlations along a $2d$ random walk, giving for example,
$C({\bf 0},z) \sim z^{-1}$.

\begin{figure}
\includegraphics[scale=0.5]{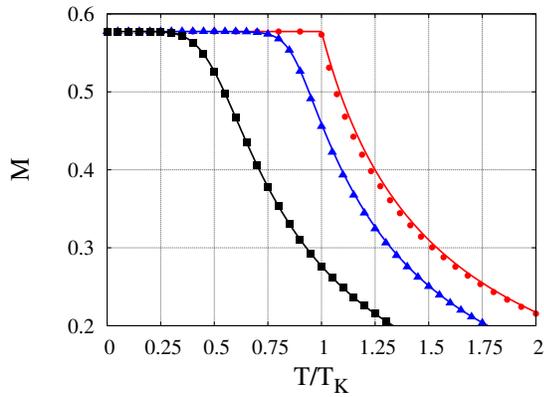}
\caption{(Color online) $M$ vs. $T/T_{\rm K}$ obtained
  from simulations for the pyrochlore lattice (dots) and analytically
  on the Bethe lattice (solid lines) for
  $h/J=10^{-3}$($\bullet$)$,\:0.13$($\blacktriangle$) and $0.58$($\blacksquare$).} \label{loop}
\end{figure}

\begin{figure}
\includegraphics[scale=0.5]{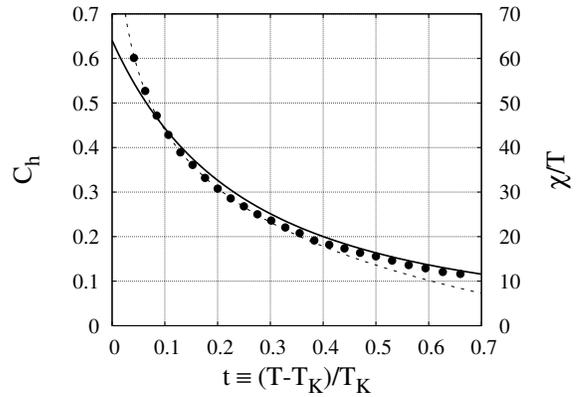}
\caption{$C_{\rm h}$ and $\chi$ vs $t \equiv \left(T/T_{\rm K}-1\right)$ obtained
  from simulations ($\bullet$) and analytically (solid lines) for $h =
  10^{-3}  J$. The dashed line is a fit to a logarithmic
  singularity derived from Eq.~(\ref{magn}).} \label{Ch-lin}
\end{figure}

We have investigated the transition via Monte Carlo (MC)
simulations, using a cluster
algorithm~\cite{Barkema98,Melko01,Isakov04PRB}. 
This allows us to avoid the apparent first-order
discontinuities~\cite{Harris98} that arise from loss of ergodicity
with single spin flip dynamics for $T\ll J$.  Each MC move consists of
the reversal of all spins on a string, which in the limit $T/
J\rightarrow 0$
and with periodic boundary conditions
must close on itself. The algorithm is efficient because in this
limit all attempted flips are accepted below saturation.  A typical
simulation involves $4 \times 10^6$ spins, studied for $10^6$ MC steps
at each temperature.
Monte Carlo data for magnetisation (Fig.~\ref{loop}), differential
susceptibility and heat capacity (Fig.~\ref{Ch-lin}) show
conclusively a 3$d$ Kasteleyn transition in the limit $h \ll J$.
For $T<T_{\rm K}$, the system is completely frozen, with $M=M_{\rm
sat}$ and $C_{\rm h} = \chi = 0$, while for $T>T_{\rm K}$, $M$
varies continuously with $T/T_{\rm K}$. 
The behaviour of $C_{\rm h}$ and $\chi$ close to $T_K$, related by $C_{\rm h} = h^2 \chi/T$, is consistent with the logarithmic singularity expected from Eq.~(\ref{magn}). Away
from the limit $h \ll J$, there is no sharp transition: instead
$M$ varies smoothly with $T$. 

\begin{figure}
\includegraphics[scale=0.5]{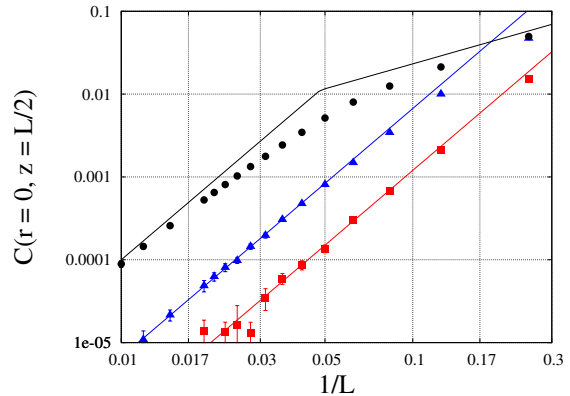}
\caption{(Color online) Finite-size scaling of spin correlations 
for $h/h_c=0.33$($\blacksquare$)$,\:0.66$($\blacktriangle$) and
  $0.93$($\bullet$), on log-log scale. Lines have slopes $3$ and $1$.}
  \label{Correlation}
\end{figure}

Data for the correlation function, illustrated in
Fig.~\ref{Correlation},  show the behaviour expected from
Eq.~(\ref{correlation-function}). At the maximum separation along
the $[100]$ direction in a system of linear size $L$ ($r{=}0$ and
$z{=}L/2$) correlations decrease with increasing $L$ as $L^{-3}$,
and increase in amplitude with $h/h_{\rm c}$. Close to the
transition, a slower decrease is observed at short distances, 
consistent with the
$L^{-1}$ law expected for isolated strings.

We supplement our simulations with analytical calculations for a
related model: a Bethe lattice (BL) of tetrahedra, whose central
element is illustrated in Fig.~\ref{pyrochlore-spin}. As on the pyrochlore lattice, each tetrahedron is connected to four others. Crucially, however, there are no closed loops of tetrahedra \cite{Chandra94}: the statistical mechanics can then be solved exactly and a Kasteleyn transition appears at $T_K$. Analytical results for the BL are very close to the
simulation data for the pyrochlore lattice -- in the absence of
adjustable parameters (Fig.~\ref{loop} and~\ref{Ch-lin}) -- providing {\it a
  posteriori} justification for the BL approach. Small differences are
apparent only for $h \ll J$: these arise because the transition on the BL is necessarily mean field, and so does not have the logarithmic contributions to
critical behaviour of Eq.~(\ref{magn}).

\begin{figure}
\includegraphics[scale=0.5]{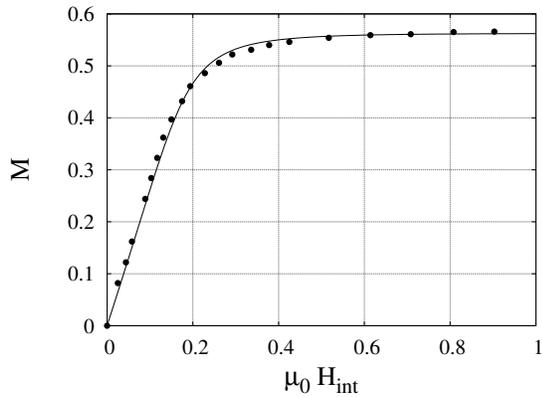}
\caption{Magnetisation $M$ vs the internal magnetic strength $\mu_0 H_{int}$ obtained analytically for the BL (solid lines) and experimentally (points) for the compound Dy$_2$Ti$_2$O$_7$ at $T_{exp}=1.8 {\rm K}$; data
from Ref.~\cite{Fukazawa02}.} \label{DTO}
\end{figure}

What about the experimental situation? 
Studies of magnetization as a function of $[100]$ field strength at
fixed low temperature have been reported for the spin ice materials
Dy$_2$Ti$_2$O$_7$ \cite{Fukazawa02} and  Ho$_2$Ti$_2$O$_7$
\cite{Fennell05}. Here we present a comparison of our theoretical results
with data for Dy$_2$Ti$_2$O$_7$.
The measured magnetisation
flattens off abruptly at the saturation value, $M_{exp}=
10\mu_B/\sqrt{3}$, in a way reminiscent of the Kasteleyn
transition. For temperatures well below the scale set by the
exchange constant, experiments indicate that spin ice materials are out of
equilibrium, as expected if the actual dynamics is local.
We therefore make a comparison at a temperature comparable with the
exchange energy. At such temperatures the behaviour of the pyrochlore
lattice and BL are indistinguishable. We show in Fig.~\ref{DTO} data for Dy$_2$Ti$_2$O$_7$ at $T_{exp}=1.8{\rm K}$ (Ref.~\cite{Fukazawa02}) together with our BL results at a fitting temperature $T=0.87\,T_{exp}$. This temperature is about $1.6$ times  the effective nearest-neighbour exchange constant $J_{eff} S^2/k_B \sim 1.11{\rm K}$ estimated for Dy$_2$Ti$_2$O$_7$~\cite{Fukazawa02}. The agreement between experiment and our theory is clearly good, indicating that the former is indeed close to the transition discussed here.
%

In conclusion, we propose that spin ice materials provide a new opportunity for the study of the Kasteleyn transition. The $3d$ transition presented here differes from that in $2d$,  predicted to occur~\cite{Moessner03} for spin ice materials in a $[111]$ field, both in its critical properties and in the behavior of correlation functions. In $2d$, it takes the form of a commensurate-incommensurate phase transition which can be evidenced by a divergent peak in the correlation function (absent in $3d$), the wavevector of which is field-dependent \cite{Moessner03}, as reported in recent measurements \cite{Fennell07}. We believe that the data of Refs.~\cite{Fukazawa02} and \cite{Fennell05} provide evidence for rounded $3d$ Kasteleyn transitions in Dy$_2$Ti$_2$O$_7$ and Ho$_2$Ti$_2$O$_7$ respectively, which could be more closely approached in future experiments.

It is a pleasure to thank S.~T. Bramwell, C. Castelnovo, T. Fennell and
M.~J. Harris for useful discussions. 
We acknowledge financial
support from the European Science Foundation PESC/RNP/HFM (PCWH and LJ), from the
Royal Society and the London Centre for
Nanotechnology (PCWH), from ANR grant 05-BLAN-0105 (LJ), and
from EPSRC Grant No. GR/83712/01 (JTC).  LJ and PCWH thank
the Rudolph Peierls Centre for Theoretical Physics, University of
Oxford, where this work was completed.

\bibliographystyle{apsrev}
\bibliography{K-bibli}

\begin{thebibliography}{25}
\expandafter\ifx\csname natexlab\endcsname\relax\def\natexlab#1{#1}\fi
\expandafter\ifx\csname bibnamefont\endcsname\relax
  \def\bibnamefont#1{#1}\fi
\expandafter\ifx\csname bibfnamefont\endcsname\relax
  \def\bibfnamefont#1{#1}\fi
\expandafter\ifx\csname citenamefont\endcsname\relax
  \def\citenamefont#1{#1}\fi
\expandafter\ifx\csname url\endcsname\relax
  \def\url#1{\texttt{#1}}\fi
\expandafter\ifx\csname urlprefix\endcsname\relax\def\urlprefix{URL }\fi
\providecommand{\bibinfo}[2]{#2}
\providecommand{\eprint}[2][]{\url{#2}}

\bibitem[{\citenamefont{Youngblood and Axe}(1981)}]{Youngblood}
\bibinfo{author}{\bibfnamefont{R.~W.} \bibnamefont{Youngblood}}
  \bibnamefont{and} \bibinfo{author}{\bibfnamefont{J.~D.} \bibnamefont{Axe}},
  \bibinfo{journal}{Phys. Rev. B {\bf 23}, 232}  (\bibinfo{year}{1981}).

\bibitem[{\citenamefont{Henley}(1997)}]{Henley}
\bibinfo{author}{\bibfnamefont{C.~L.} \bibnamefont{Henley}},
  \bibinfo{journal}{J. Stat Phys. {\bf 89}, 483}  (\bibinfo{year}{1997}).

\bibitem[{\citenamefont{Chakraborty et~al.}(2002)\citenamefont{Chakraborty,
  Das, and Kondev}}]{Kondev}
\bibinfo{author}{\bibfnamefont{B.}~\bibnamefont{Chakraborty}},
  \bibinfo{author}{\bibfnamefont{D.}~\bibnamefont{Das}}, \bibnamefont{and}
  \bibinfo{author}{\bibfnamefont{J.}~\bibnamefont{Kondev}},
  \bibinfo{journal}{Eur. Phys. J. E {\bf 9}, 227}  (\bibinfo{year}{2002}).

\bibitem[{\citenamefont{Anderson}(1956)}]{Anderson56}
\bibinfo{author}{\bibfnamefont{P.~W.} \bibnamefont{Anderson}},
  \bibinfo{journal}{Phys. Rev. {\bf 102}, 1008}  (\bibinfo{year}{1956}).

\bibitem[{\citenamefont{Harris~{\it et al.}}(1997)}]{Harris97}
\bibinfo{author}{\bibfnamefont{M.~J.} \bibnamefont{Harris~{\it et al.}}},
  \bibinfo{journal}{Phys. Rev. Lett. {\bf 79}, 2554}  (\bibinfo{year}{1997}).

\bibitem[{\citenamefont{Moessner and Chalker}(1998)}]{Moessner98}
\bibinfo{author}{\bibfnamefont{R.}~\bibnamefont{Moessner}} \bibnamefont{and}
  \bibinfo{author}{\bibfnamefont{J.~T.} \bibnamefont{Chalker}},
  \bibinfo{journal}{Phys. Rev. B {\bf 58}, 12049}  (\bibinfo{year}{1998}).

\bibitem[{\citenamefont{Ramirez~{\it et al.}}(1999)}]{Ramirez99}
\bibinfo{author}{\bibfnamefont{A.~P.} \bibnamefont{Ramirez~{\it et al.}}},
  \bibinfo{journal}{Nature {\bf 399}, 333}  (\bibinfo{year}{1999}).

\bibitem[{\citenamefont{Bhattacharjee~{\it et al.}}(1983)}]{Bhattacharjee83}
\bibinfo{author}{\bibfnamefont{S.~M.} \bibnamefont{Bhattacharjee~{\it et
  al.}}}, \bibinfo{journal}{J. Stat. Phys. {\bf 32}, 361}
  (\bibinfo{year}{1983}).

\bibitem[{\citenamefont{Kasteleyn}(1963)}]{Kasteleyn63}
\bibinfo{author}{\bibfnamefont{P.~W.} \bibnamefont{Kasteleyn}},
  \bibinfo{journal}{J. Math. Phys. {\bf 4}, 287}  (\bibinfo{year}{1963}).

\bibitem[{\citenamefont{Nagle}(1973)}]{Nagle73}
\bibinfo{author}{\bibfnamefont{J.~F.} \bibnamefont{Nagle}},
  \bibinfo{journal}{Proc. Nat. Acad. Sci. USA {\bf 70}, 3443}
  (\bibinfo{year}{1973}).

\bibitem[{\citenamefont{Harris~{\it et al.}}(1998)}]{Harris98}
\bibinfo{author}{\bibfnamefont{M.~J.} \bibnamefont{Harris~{\it et al.}}},
  \bibinfo{journal}{Phys. Rev. Lett. {\bf 81}, 4496}  (\bibinfo{year}{1998}).

\bibitem[{\citenamefont{Fukazawa~{\it et al.}}(2002)}]{Fukazawa02}
\bibinfo{author}{\bibfnamefont{H.}~\bibnamefont{Fukazawa~{\it et al.}}},
  \bibinfo{journal}{Phys. Rev. B {\bf 65}, 054410}  (\bibinfo{year}{2002}).

\bibitem[{\citenamefont{Siddharthan~{\it et al.}}(1999)}]{Siddharthan99}
\bibinfo{author}{\bibfnamefont{R.}~\bibnamefont{Siddharthan~{\it et al.}}},
  \bibinfo{journal}{Phys. Rev. Lett. {\bf 83}, 1854}  (\bibinfo{year}{1999}).

\bibitem[{\citenamefont{den Hertog and Gingras}(2000)}]{Hertog00}
\bibinfo{author}{\bibfnamefont{B.~C.} \bibnamefont{den Hertog}}
  \bibnamefont{and} \bibinfo{author}{\bibfnamefont{M.~J.~P.}
  \bibnamefont{Gingras}}, \bibinfo{journal}{Phys. Rev. Lett. {\bf 84}, 3430}
  (\bibinfo{year}{2000}).

\bibitem[{\citenamefont{Isakov et~al.}(2005)\citenamefont{Isakov, Moessner, and
  Sondhi}}]{Isakov05}
\bibinfo{author}{\bibfnamefont{S.~V.} \bibnamefont{Isakov}},
  \bibinfo{author}{\bibfnamefont{R.}~\bibnamefont{Moessner}}, \bibnamefont{and}
  \bibinfo{author}{\bibfnamefont{S.~L.} \bibnamefont{Sondhi}},
  \bibinfo{journal}{Phys. Rev. Lett. {\bf 95}, 217201}  (\bibinfo{year}{2005}).

\bibitem[{\citenamefont{Fisher and Hohenberg}(1988)}]{Fisher88}
\bibinfo{author}{\bibfnamefont{D.~S.} \bibnamefont{Fisher}} \bibnamefont{and}
  \bibinfo{author}{\bibfnamefont{P.~C.} \bibnamefont{Hohenberg}},
  \bibinfo{journal}{Phys. Rev. B {\bf 37}, 4936}  (\bibinfo{year}{1988}).

\bibitem[{\citenamefont{Nagle et~al.}(1989)\citenamefont{Nagle, Yokoi, and
  Bhattacharjee}}]{Nagle89}
\bibinfo{author}{\bibfnamefont{J.~F.} \bibnamefont{Nagle}},
  \bibinfo{author}{\bibfnamefont{C.~S.~O.} \bibnamefont{Yokoi}},
  \bibnamefont{and} \bibinfo{author}{\bibfnamefont{S.~M.}
  \bibnamefont{Bhattacharjee}}, \emph{\bibinfo{title}{Phase Transitions and
  Critical Phenomena, Ed. C. Domb and J. L. Lebowitz}}
  (\bibinfo{publisher}{Academic Press}, \bibinfo{year}{1989}),
  chap.~\bibinfo{chapter}{2}.

\bibitem[{\citenamefont{Isakov~{\it et al.}}(2004{\natexlab{a}})}]{Isakov04}
\bibinfo{author}{\bibfnamefont{S.~V.} \bibnamefont{Isakov~{\it et al.}}},
  \bibinfo{journal}{Phys. Rev. Lett. {\bf 93}, 167204}
  (\bibinfo{year}{2004}{\natexlab{a}}).

\bibitem[{\citenamefont{Barkema and Newman}(1998)}]{Barkema98}
\bibinfo{author}{\bibfnamefont{G.~T.} \bibnamefont{Barkema}} \bibnamefont{and}
  \bibinfo{author}{\bibfnamefont{M.~E.~J.} \bibnamefont{Newman}},
  \bibinfo{journal}{Phys. Rev. E {\bf 57}, 1155}  (\bibinfo{year}{1998}).

\bibitem[{\citenamefont{Melko et~al.}(2001)\citenamefont{Melko, den Hertog, and
  Gingras}}]{Melko01}
\bibinfo{author}{\bibfnamefont{R.~G.} \bibnamefont{Melko}},
  \bibinfo{author}{\bibfnamefont{B.~C.} \bibnamefont{den Hertog}},
  \bibnamefont{and} \bibinfo{author}{\bibfnamefont{M.~J.~P.}
  \bibnamefont{Gingras}}, \bibinfo{journal}{Phys. Rev. Lett. {\bf 87}, 067203}
  (\bibinfo{year}{2001}).

\bibitem[{\citenamefont{Isakov~{\it et al.}}(2004{\natexlab{b}})}]{Isakov04PRB}
\bibinfo{author}{\bibfnamefont{S.~V.} \bibnamefont{Isakov~{\it et al.}}},
  \bibinfo{journal}{Phys. Rev. B {\bf 70}, 104418}
  (\bibinfo{year}{2004}{\natexlab{b}}).

\bibitem[{\citenamefont{Chandra and Doucot}(1994)}]{Chandra94}
\bibinfo{author}{\bibfnamefont{P.}~\bibnamefont{Chandra}} \bibnamefont{and}
  \bibinfo{author}{\bibfnamefont{B.}~\bibnamefont{Doucot}},
  \bibinfo{journal}{J. Phys. A {\bf 27}, 1541}  (\bibinfo{year}{1994}).

\bibitem[{\citenamefont{Fennell~{\it et al.}}(2005)}]{Fennell05}
\bibinfo{author}{\bibfnamefont{T.}~\bibnamefont{Fennell~{\it et al.}}},
  \bibinfo{journal}{Phys. Rev. B {\bf 72}, 224411}  (\bibinfo{year}{2005}).

\bibitem[{\citenamefont{Moessner and Sondhi}(2003)}]{Moessner03}
\bibinfo{author}{\bibfnamefont{R.}~\bibnamefont{Moessner}} \bibnamefont{and}
  \bibinfo{author}{\bibfnamefont{S.~L.} \bibnamefont{Sondhi}},
  \bibinfo{journal}{Phys. Rev. B {\bf 68}, 064411}  (\bibinfo{year}{2003}).

\bibitem[{\citenamefont{Fennell~{\it et al.}}(2007)}]{Fennell07}
\bibinfo{author}{\bibfnamefont{T.}~\bibnamefont{Fennell~{\it et al.}}},
  \bibinfo{journal}{Nature Physics {\bf 3}, 566}  (\bibinfo{year}{2007}).

\end{thebibliography}


\end{document}